
\baselineskip=2\baselineskip
\magnification=\magstep1
\def\pui{10^{10}}
\def\puis{10^{-5}}
\def\li{^7 Li}
\def\he{^3 He}
\def\hel{^4 He}

\def\et{\eta _{10}}
\def\rat{D/H}
\def\omen{\Omega _N}
\centerline{\bf COSMIC DEUTERIUM OR A HYDROGEN INTERLOPER?}
\vskip12pt
\centerline{\bf Gary Steigman}
\centerline{\bf Institute of Astronomy, Madingley Road, Cambridge CB3 0HA}
\vskip36pt
\centerline{\bf Summary}
\vskip6pt
Two groups have independently reported the possible detection of extragalactic
deuterium in the absorption spectrum of the same high redshift, low metalicity
QSO.  Although the high value for the inferred deuterium abundance poses no
problems for cosmology (i.e., big bang nucleosynthesis), it is in apparent
conflict with solar system observations of deuterium and helium-3.  This latter
inconsistency is explained and made quantitative and it is shown that, unless
the inferred D/H ratio is too high by a factor of three, these data challenge
our understanding of the stellar and galactic evolution of helium-3.  This
conflict is resolved if the observed absorption feature is, in fact, due to a
hydrogen interloper rather than to high z, low Z deuterium.
\vskip36pt
The observed expansion of the Universe and the presence of the cosmic
background
radiation ensure that, during its early evolution, the Universe passed through
a hot, dense epoch when nuclear reactions synthesized the light nuclides
(deuterium,
helium-3, helium-4 and lithium-7).  In the context of standard Big Bang
Nucleosynthesis
(BBN), deuterium $(D)$ and helium-3 $(\he)$ are rapidly burned to helium-4
$(\hel)$
and, the higher the nucleon density (measured by $\eta$, the present ratio of
nucleons to photons;  $\et \equiv \pui \eta$), the more rapidly is $D$ (and
$\he$)
consumed.  Thus, the cosmic abundance of deuterium provides a sensitive probe
of the universal density of nucleons (Reeves et al. 1973).  Of the light
elements, $D$ is unique in
that BBN is the only astrophysical site for its production in amounts
comparable
to those observed (Epstein, Lattimer \& Schramm 1976). But, whenever
interstellar gas is incorporated in stars, $D$
is destroyed by nuclear burning.  As a result, a lower bound to the cosmic
abundance of $D$ provides a lower bound to its primordial abundance from which
we may infer an upper bound to $\eta$ (Reeves et al. 1973).
\vskip6pt
In the course of Galactic evolution, as more gas passes through new generations
of stars, $D$ is burned to $\he$.  However, not all pre-stellar $\he$ is
consumed,
with the result that observations of solar and/or interstellar $D$ and $\he$
can
provide an upper bound to the primordial (or, pre-Galactic) abundance of D
(Rood, Steigman \& Tinsley 1976 ; Dearborn, Schramm \& Steigman 1986 ; Yang et
al. 1984).
Such an upper bound to primordial $D$ leads to a lower bound to $\eta$.  Of
course, solar and interstellar material have been processed through stars in
the course of
Galactic evolution so that observations of pre-Galactic $D$ would be of great
value.  Just such - possible - observations of cosmic deuterium in a high
redshift, low metallicity QSO absorption system have recently been reported by
two independent groups (Songaila et al. 1994 ; Carswell et al. 1994).  Although
using different telescopes and detectors to
observe the $z_{abs} = 3.32$ absorption system towards the $z_{em} = 3.42$
quasar 0014 + 813, both collaborations derive a deuterium abundance ($D$ to
$H$ (hydrogen) ratio by number) of $(D/H) _{OBS} \approx 25 \times \puis$.
However, it has been noted$^7$ that there is significant probability that what
has
been identified as a deuterium feature could, in fact, be a low column density
interloper (a $Ly\alpha$ system at a velocity shift of 80 kms$^{-1}$ with
respect
to the main absorber).  It is my goal here to show that this latter possibility
is most likely correct by demonstrating the incompatibility of such a large
$D/H$ ratio with solar system observations of $D$ and $\he$.  Such an approach
has the virtue of avoiding any prejudice concerning BBN (standard,
inhomogeneous  or otherwise).
Nonetheless, the cosmological implications of these observations, if they
really
were due to $D$, are interesting and are discussed, briefly, next.
\vskip6pt
Assume for the moment that D has been detected in a nearly primordial (high
redshift, low metallicity) system.  The inferred $D/H$ ratio should place a
lower bound on the primordial abundance;  since $D$ is only destroyed
subsequent
to BBN, $(\rat)_{BBN} \geq (\rat)_{OBS}$.  For $(\rat)_{BBN} \geq 25\times
\puis,
\ \et \leq 1.5$ (Walker et al. 1991) and the corresponding contribution of
nucleons to the overall
density of the Universe (as measured by the nucleon density parameter
$\omen$ and the present value of the Hubble parameter $h_{50} \equiv H_0
/50kms^{-1}
Mpc^{-1}$) is limited to:  $\omen h^2 _{50} \leq 0.022$.  Even for $H_0 \geq 40
kms^{-1}Mpc^{-1}$, this places a severe upper bound on the nucleon density
$(\omen \leq 0.034)$, strengthening the BBN case for non-baryonic dark matter.
What of the BBN abundances of the other light elements?  With decreasing
nucleon
density the primordial yield of $\hel$ decreases and that of $\li$ increases.
For $\et \approx 1.5$, $(\li /H)_{BBN} \approx 2.3$ which is comparable to the
upper bound inferred from observations of lithium in the most metal-poor Pop II
stars ( Walker et al. 1991). And, such a low value of $\eta$ actually improves
the consistency
between the BBN predicted yield of $\hel$ and the primordial abundance
inferred from observations
of very low metallicity, extragalactic HII regions.  The observations suggest
a primordial mass fraction of $\hel$, $Y = 0.23 \pm 0.01$ (Olive, Steigman \&
Walker 1991) while,
for $\et \approx 1.5$, $Y_{BBN} \approx 0.231$.  Thus, the low value of $\eta$
implied by a large primordial abundance of $D$ is entirely compatible with the
observed abundances and the BBN predicted yields of $\hel$ and $\li$ (and,
$\he$
as well).
\vskip6pt
However, as I shall now show, it is very likely that extragalactic deuterium
has not yet been detected.  The problem lies in the comparison between the very
large abundance inferred from the QSO observations $\left( (\rat)_{OBS} \approx
25 \times \puis \right)$ and the much smaller solar system and interstellar
medium abundances (Walker et al. 1991 ; Geiss 1994) ($(\rat)_\odot = 2.6 \pm
0.9 \times \puis$;
$(\rat)_{ISM} = 1.6 \pm 0.2 \times \puis$).  Consistency among these data would
require large destruction of $D$ in the course of Galactic evolution.  But,
since
$D$ is burned in stars to $\he$ and some $\he$ survives stellar processing,
this
would imply an enhanced abundance of $\he$.  To make this quantitative, let us
concentrate on the solar system where meteoritic, lunar and solar wind data
provide statistically accurate estimates of both the $D$ and $\he$ abundances
in the pre-solar nebula (Geiss 1994).
\vskip6pt
Neglecting any net stellar production of $\he$ as well as any contribution
from primordial $\he$
(to maximize our bound), Yang et al.(1984) derived the following inequality
which we
apply to the solar system,
$$
{D + \he \overwithdelims () H}_{BBN} \leq {D \overwithdelims () H}_\odot +
{1 \over g_3}{\he \overwithdelims () H}_\odot . \eqno{(1)}
$$
In equation (1) there are no assumptions concerning BBN and, the only
dependence on Galactic evolution enters through $g_3$, the stellar survival
fraction of $\he$(Dearborn, Schramm \& Steigman 1986). Equation (1) is a simple
reflection of the result that, since
$D$ burns to $\he$, the observed abundances of $D$ and $\he$ provide an upper
bound to the sum of the primordial abundances of $D$ plus $\he$.  Since, here,
we are
interested in bounding primordial $D$ and, $\rat \leq (D + \he)/H$,
$$
(\rat)_{BBN} \leq (\rat)_\odot + g_{3} ^{-1} \left(He/H \right)_\odot .
\eqno{(2)}
$$
The stellar models of Dearborn, Schramm and Steigman (1986) suggested that $g_3
\ge 1/4$ and this is confirmed by the galactic evolution calculations of
Steigman and Tosi (1992) who, for a range of different models, found $1/3 \leq
g_{3}
\leq 1/2$.  Using the Geiss (1994) solar system abundances and $g_3 \geq 1/4$,
we
derive an upper bound to the BBN abundance of $D$.
$$
10^5 (\rat)_{BBN} \leq 8.6 \pm 1.5 \eqno{(3)}
$$

Here lies the problem with the interpretation of the observed QSO absorption
feature as
due to $D$.  If correct, we should have $(\rat )_{OBS} \leq (\rat)_{BBN}$.
Instead, we find that $(\rat)_{OBS}$ exceeds the upper bound on $(\rat)_{BBN}$
by nearly a factor of three.  Put another way, the solar system bound on the
BBN abundance is smaller than the inferred abundance in the QSO absorption
system by some 12 sigma.  The large value of $(D/H)_{OBS}$ could only be
compatible with the low solar system abundances of $D$ and $\he$ if the
$\he$ survival fraction has been significantly overestimated (Dearborn,
Schramm \& Steigman 1986 ; Steigman \& Tosi 1992);
consistency is only obtained for $g_3 \leq 0.09 \pm 0.02$, which is some 8
sigma away from the standard lower bound (Dearborn, Schramm \& Steigman 1986)
of $g_3 \geq 0.25$.
\vskip6pt
If, indeed, this absorption feature is not due to intergalactic deuterium, what
is it?  As Carswell et al. (1994) have noted, the probability that any single
measurement is confused with hydrogen absorption is high $(\sim 15 \%)$.  The
observed feature could be from hydrogen absorption in a low column density
cloud along the same line of sight, displaced in velocity from the main
absorber by 80 kms$^{-1}$.  Thus, any such
observation can only place an upper bound on the pre-Galactic $D$-abundance.
Although it is likely that nearly primordial $D$ is yet to be observed,  the
capability of current telescopes and detectors to achieve such an important
observation is clear.  The observers are to be commended and encouraged to
press
on.

\vskip12pt

\noindent References

\item{1.} Reeves, H., Audouze, J., Fowler, W.A. and Schramm, D.N.
{\it Ap.J.} {\bf 179}, 909--930 (1973).
\item{2.} Epstein, R.I., Lattimer, J.M. and Schramm, D.N.
{\it Nature}, {\bf 263}, 198--202 (1976).
\item{3.} Rood, R.T., Steigman, G. and Tinsley, B.M. {\it Ap. J.}
{\bf 207}, L57--L60 (1976).
\item{4.} Dearborn, D.S.P., Schramm, D.N. and Steigman, G. {\it Ap. J.}
{\bf 302}, 35--38 (1986).
\item{5.} Yang, J., Turner, M.S., Steigman, G., Schramm, D.N. and Olive, K.A.
{\it Ap. J.} {\bf 281}, 493--511 (1984).
\item{6.} Songaila, A., Cowie, L.L., Hogan, C. and Rugers, M. {\it Nature},
In Press (1994).
\item{7.} Carswell, R.F., Rauch, M., Weymann, R.J., Cooke, A.J. and Webb, J.K.
{\it MNRAS}, In Press (1994).
\item{8.} Walker, T.P., Steigman, G., Schramm, D.N., Olive, K.A. and Kang, H.S.
{\it Ap. J.} {\bf 376}, 51--69 (1991).
\item{9.} Olive, K.A., Steigman, G. and Walker, T.P. {\it Ap. J.} {\bf 380},
L1--L4 (1991).
\item{10.} Geiss, J. In Press, {\it Symposium on the Origin and Evolution of
the
Elements} (eds., N. Prantzos, E. Flam and M. Cass\'e; Cambridge University
Press)
(1994).
\item{11.} Steigman, G. and Tosi, M. {\it Ap. J.} {\bf 401}, 150--156 (1992).
\bye